\documentclass[twocolumn,aps,prd,amssymb,eqsecnum,nofootinbib]{revtex4}
\usepackage{amsmath}
\usepackage{amssymb}
\usepackage{epsfig}
\usepackage{bm}
\begin{document}

\title{Post-Newtonian Approximation in Maxwell-Like Form}

\author{Jeffrey D.\ Kaplan, David A.\ Nichols, and Kip S. Thorne }
\affiliation{Theoretical Astrophysics, California Institute of Technology,
Pasadena, CA 91125.}
\date{\today}

\begin{abstract}
The equations of the {\it linearized} first post-Newtonian approximation to general relativity are 
often written in ``gravitoelectromagnetic'' Maxwell-like form, since that facilitates physical
intuition.  Damour, Soffel and Xu (DSX) (as a side issue in their complex but elegant papers on
relativistic celestial mechanics) have expressed the first post-Newtonian approximation, {\it including all nonlinearities},
in Maxwell-like form.  This paper summarizes that DSX Maxwell-like formalism (which is not easily extracted
from their celestial mechanics papers), and then extends it to include the post-Newtonian (Landau-Lifshitz-based) gravitational
momentum density, momentum flux (i.e.\ gravitational stress tensor) and law of momentum
conservation in Maxwell-like form.  The authors and their colleagues have found these Maxwell-like
momentum tools useful 
for developing physical intuition into numerical-relativity simulations of compact binaries with spin.

\end{abstract}


\maketitle

\section{Introduction}
\label{sec:intro}

In 1961,
Robert L.\ Forward \cite{Forward} 
(building on earlier work of Einstein \cite{Einstein1,Einstein2} and especially Thirring \cite{Thirring1,Thirring2}) 
wrote the \textit{linearized, slow-motion approximation  to general relativity} in a form that closely resembles Maxwell's equations; and he displayed 
this formulation's  great intuitive and computational power.
In the half century since then, this Maxwell-like formulation and
variants of it have been widely explored and used; see, e.g., \cite{BCT, Wald, Mashhoon93, MashhoonGonwaldLichtenegger, ClarkTucker,Mashhoon08,GoulartFalciano,CostaHerdeiro} and references therein.  

In 1965--69 S.\ Chandrasekhar \cite{ChandraPN,ChandraPNcons} formulated the \textit{first post-Newtonian (weak-gravity, slow-motion) approximation to general relativity} in a manner that has been widely used for astrophysical calculations during the subsequent 40 years.
When linearized, this first post-Newtonian (1PN) approximation can be (and often is) recast in Maxwell-like form.  

In 1991, T.\ Damour, M.\ Soffel and C.\ Xu (DSX; \cite{DSX1}) extended this Maxwell-like 1PN formalism to include all
1PN nonlinearities (see also Sec.\ 13 of Jantzen, Carini and Bini \cite{JCB}).  DSX did so as a tool in developing a general formalism
for the celestial mechanics of bodies that have arbitrary internal structures and correspondingly
have external gravitational fields
characterized by two infinite sets of multipole moments.  (For a generalization to scalar-tensor theories,
see \cite{Kopeikin}.)  In 2004, Racine and Flanagan \cite{RacineFlanagan} generalized DSX to a system of compact bodies (e.g.\ black holes) that have arbitrarily strong internal gravity.  

During the past 18 months, we and our colleagues have been exploring the flow of gravitational field momentum in
numerical-relativity simulations of compact, spinning binaries \cite{CKNT,Lovelace}.  In their inspiral phase, these binaries' 
motions and precessions can be described by the 1PN approximation,\footnote{For black-hole and 
neutron-star binaries, the influences of spin that interest us are formally 1PN, but because of the bodies' 
compactness (size of order Schwarzschild radius), they are numerically 1.5PN.}  and we have gained much insight
into their dynamics by using the 1PN DSX Maxwell-like formalism, {\it extended}
to include 
Maxwell-like momentum density, momentum flux, and momentum conservation.\footnote{A referee has
pointed out to us that some papers in the rich literature on the Maxwell-like formulation of linearized 1PN theory, e.g.\ \cite{ClarkTucker}, argue that the Maxwell analogy is physically useful only for stationary phenomena.  Our spinning-binary application \cite{CKNT} of the momentum-generalized DSX formalism 
is a counterexample.}

In this paper, we present that extension of DSX,\footnote{When we carried out our analysis
and wrote it up in the original version of this paper,
we were unaware of the Maxwell-like formalism in
DSX \cite{DSX1}; see our preprint  at \url{http://xxx.lanl.gov/abs/0808.2510v1} When we learned of DSX from Luc Blanchet, we used it to improve our Maxwell-like treatment of gravitational momentum (by
replacing our definition for the gravitoelectric field by that of DSX) and 
we rewrote this paper to highlight the connection to DSX.}  
though with two specializations:
(i) we fix our coordinates (gauge)
to be fully harmonic instead of maintaining the partial gauge invariance of DSX, and (ii) we discard all
multipole moments of the binaries' bodies except their masses and their spin angular momenta, because
for black holes and neutron stars, the influences of all other moments are numerically much smaller
than 1.5PN order.  

The DSX celestial-mechanics papers \cite{DSX1,DSX2,DSX3,DSX4} are so long and complex that it is not easy to
extract from them the bare essentials of the DSX Maxwell-like 1PN formalism.  For the benefit of researchers who want
those bare essentials and want to see how they are related to more conventional
approaches to 1PN theory, we summarize them before presenting our momentum extension, and we 
do so for a general stress-energy tensor, for a 
perfect fluid, and for a system of compact bodies described by their masses and spins.

This paper is organized as follows:  In Sec.\ \ref{sec:DSX} we summarize the basic DSX
equations for 1PN theory in Maxwell-like form.  
In Sec.\ \ref{sec:PerfectFluid} 
we specialize the DSX formalism to a perfect fluid and make contact with the conventional 1PN notation.
In Sec.\ 
\ref{sec:Momentum} we extend DSX by deriving the (Landau-Lifshitz-based) density and flux of gravitational momentum in terms
of the DSX gravitoelectric and gravitomagnetic fields and by writing down the law 
of momentum conservation in terms of them.  (It is this that we have found so useful for gaining intuition
into numerical-relativity simulations of inspiraling, spinning binaries \cite{CKNT,Lovelace}.)  In Sec.\ \ref{sec:energyconservation} we briefly discuss energy conservation.
In Sec.\ \ref{sec:vacuum}, relying on Racine and Flanagan \cite{RacineFlanagan}, we specialize to the vacuum in the near zone of a system made
from compact bodies with arbitrarily strong internal gravity. 
Finally, in Sec.\ \ref{sec:conclusion} we summarize the DSX formalism and our extension of it
both for
a self-gravitating fluid and for a system of compact bodies.

Throughout this paper, we set $G=c=1$, Greek 
letters run from 0 to 3 (spacetime) and Latin from 1 to 3 (space), and we use the notation of field theory
in flat space in a 3+1 split, so spatial indices are placed up or down equivalently and repeated
spatial indices are summed whether up or down or mixed.  We use bold-face italic characters to
represent spatial vectors, i.e.\ $\bm w$ is the bold-face version of $w_j$.

\section{The DSX Maxwell-Like Formulation of 1PN Theory}
\label{sec:DSX}

Damour, Soffel and Xu (DSX \cite{DSX1}) express the 1PN metric in terms of two gravitational
potentials, a scalar $w$ and a vector $w_j$:
\begin{eqnarray}
g_{00} &=& -e^{-2w} = -1 + 2w - 2w^2 + O(U_N^3)\;, \nonumber \\
g_{0i} &=& -4 w_i + O(U_N^{5/2})\;,
\label{eqn:DSXmetric} \\
g_{ij} &=& \delta_{ij} e^{2w} = \delta_{ij} (1+2w) + O(U_N^2)\nonumber
\end{eqnarray}
[DSX Eqs.\ (3.3)].  The Newtonian limit of $w$ is $U_N = ($Newtonian gravitational potential),
and $w_i$ is of order $U_N^{3/2}$:  
\begin{equation}
w = U_N + O(U_N^2)\;, \quad w_i = O(U_N^{3/2})\;.
\label{eqn:wMag}
\end{equation}

The harmonic gauge condition implies that
\begin{equation}
w_{,t} + w_{j,j} = 0
\label{eqn:DSXharmonic}
\end{equation}
[DSX Eq.\ (3.17a)]; here and throughout commas denote partial derivatives.  Using this
gauge condition (which DSX do not impose), the 1PN Einstein field equations take the
following remarkably simple form:
\begin{subequations}
\begin{eqnarray}
\nabla^2 w - \ddot{w} &=& -4\pi (T^{00} + T^{jj}) + O(U_N^3/{\mathcal L}^2)\;, 
\label{eqn:EFEw}
\\
\nabla^2 w_i &=& - 4\pi T^{0i} + O(U_N^{5/2}/{\mathcal L}^2)
\label{eqn:EFEwi}
\end{eqnarray}
\label{eqn:DSXEFE}
\end{subequations}
[DSX Eqs.\ (3.11)].
Here $T^{\alpha\beta}$ is the stress-energy tensor of the source (which we specialize below to 
a perfect fluid), $\nabla^2$ is the flat-space Laplacian (i.e.\ $\nabla^2 w = w_{,jj}$), repeated indices are summed, dots denote time derivatives
(i.e.\ $\ddot{w} = w_{,tt}$), and $\mathcal L$ is the lengthscale on which $w$ varies.

Following DSX, we introduce the 1PN \textit{gravitoelectric field} $\bm g$ (denoted $\bm e$
or $\bm E$ by DSX , depending on the context) and \textit{gravitomagnetic field} $\bm H$ 
(denoted $\bm b$ or $\bm B$ by DSX):
\begin{subequations}
\begin{eqnarray}
\bm g &=& \boldsymbol \nabla w + 4 \dot{\bm w} + O(U_N^3/{\mathcal L})\;, 
\label{eqn:DSXgDef} \\
\bm H &=&- 4 \boldsymbol \nabla \times \bm w + O(U_N^{5/2}/\mathcal L)\;.
\label{eqn:DSXHDef}
\end{eqnarray}
\label{eqn:DSXghDef}
\end{subequations}
[DSX Eqs.\ (3.21)].  

The Einstein equations (\ref{eqn:DSXEFE}) and these definitions imply the following
1PN Maxwell-like equations for $\bm g$ and $\bm H$:
\begin{subequations}
\begin{eqnarray}
\bm \nabla \cdot \bm g &=& -4\pi(T^{00} + T^{jj}) - 3 \ddot w + O( g U_N^2/\mathcal L)\;, 
\label{eqn:divg} \\
\bm \nabla \times \bm g &=& - \dot{\bm H}+ O(g U_N^2 /\mathcal L)\;,
\label{eqn:curlg} \\
\bm \nabla \cdot \bm H &=& 0 + O(H U_N /\mathcal L)\;,
\label{eqn:divH} \\
\bm \nabla \times \bm H &=& - 16\pi T^{0i} \bm e_i + 4 \dot{\bm g} + O(H U_N /\mathcal L)
\label{eqn:curlH} 
\end{eqnarray}
\label{eqn:gMaxwell}
\end{subequations}
[DSX Eqs. (3.22)]. Here $\bm e_i$ is the unit vector in
the $i$ direction.

In terms of $\bm g$ and $\bm H$, the geodesic equation for a particle with ordinary velocity $\bm v = d\bm x/dt$
takes the following form [Eq. (7.17) of DSX, though in a less transparently
``Lorentz-force''-like form there]:  
\begin{eqnarray}
\frac{d}{dt} \left[ \left(1+3U_N+\tfrac12 \bm v^2\right) \bm v\right] &=& \left(1-U_N + \tfrac32 \bm v^2\right) \bm g
+ \bm v \times \bm H 
\nonumber \\
&&+ O(g U_N^2 )\;.
\label{eqn:geodesic}
\end{eqnarray}

Note that the spatial part of the particle's 4-momentum is $m \bm u = m(1+U_N+\tfrac12 \bm v^2)\bm v$ at 1PN order.  This accounts for the coefficient $1+U_N+\tfrac12 \bm v^2$ on the
left-hand side of Eq.\ (\ref{eqn:geodesic}).  The remaining factor $2U_N$ is related to the
difference between physical lengths and times, and proper lengths and times.  In the 
linearized, very-low-velocity approximation, this geodesic equation takes the ``Lorentz-force'' form
$d\bm v/dt = \bm g + \bm v \times \bm H$, first deduced (so far as we know) in 1918 by Thirring 
\cite{Thirring1}, motivated by Einstein's 1913 \cite{Einstein1} insights about similarities between 
electromagnetic theory and his not-yet-perfected general relativity theory. 

The 1PN deviations of the geodesic equation (\ref{eqn:geodesic}) from the usual Lorentz-force form might make one
wonder about the efficacy of the DSX definition of $\bm g$.  That efficacy will show up most
strongly when we explore the gravitational momentum density in Sec.\ \ref{sec:Momentum} below.

\section{Specialization to a Perfect Fluid}
\label{sec:PerfectFluid}

We now depart from DSX by specializing our source to a perfect fluid and making contact with a set of 1PN gravitational
potentials that are widely used.  We pay special attention to connections with a paper by
Pati and Will \cite{PatiWill1} because that paper will be our foundation, in Sec.\ 
\ref{sec:Momentum}, for computing the density and flux of
gravitational field momentum.

We describe our perfect fluid in the following standard notation:
$\rho_o = ($density of rest mass),  $\Pi =$(internal energy per unit
rest mass, i.e.\ specific internal energy), $P=($pressure), all as measured in the fluid's local rest frame; 
$v_j \equiv dx^j/dt =($fluid's coordinate velocity).  

Following Blanchet and Damour \cite{BlanchetDamour}, and subsequently  Pati and Will (Eqs.\ (4.13), (4.3) of \cite{PatiWill1}), we introduce a post-Newtonian variant $U$ of the Newtonian potential,
which is sourced by $T^{00}+ T^{jj}$:  $\nabla^2 U = - 4\pi(T^{00}+T^{jj})$.  Accurate to 1PN order, the source is [see Eq.\ (\ref{eqn:T00+Tjj})]
\begin{equation}
T^{00}+ T^{jj} = \rho_o(1+\Pi+2{ \bm v}^2 + 2U_N+3P/\rho_o)\;,
\label{eqn:T00TjjFluid}
\end{equation}
where $U_N$ is the Newtonian limit of $U$
\begin{subequations}
\begin{equation}
U_N ({\bm x},t) = \int \frac{ \rho_o({\bm x}',t)}{|{\bm x} - {\bm x}'|} d^3 x'\;.
\label{eqn:UN}
\end{equation}
Correspondingly,
$U$ can be written as
\begin{equation}
U = \int \frac{\rho_o(1+\Pi+2{ \bm v}^2 + 2U_N+3P/\rho_o)} {|{\bm x} - {\bm x}'|} d^3x'\;.
\label{eqn:U}
\end{equation}
\label{eqn:UNU}
\end{subequations}
Here and below the fluid variables and gravitational potentials in the integrand are functions of $({\bm x}',t)$ as in Eq.\ (\ref{eqn:UN}).  In Eq.\ (\ref{eqn:UN}) for $U_N$, $\rho_o$ can be replaced by any quantity that agrees with $\rho_o$
in the Newtonian limit, e.g.\ by the post-Newtonian ``conserved mass density'' $\rho_*$ of Eq.\ (\ref{eqn:rho*}) below.
We also introduce Chandrasekhar's Post-Newtonian scalar gravitational potential $\chi$ (Eq.\ (44) of \cite{ChandraPN}), which is sourced by $2 U_N$, 
$\nabla^2 \chi = - 2U_N$ or equivalently
\begin{equation}
\chi = - \int \rho_o |{\bm x} - {\bm x'}| d^3 x'\;.
\label{eqn:chi}
\end{equation}
Pati and Will use the notation $-X$ for $\chi$ (Eqs.\ (4.14), (4.12a) and (4.3) of \cite{PatiWill1}).  

It is straightforward to show that the 1PN solution to the wave equation (\ref{eqn:EFEw}) for the DSX scalar
potential $w$ is
\begin{equation}
w = U - \frac12 \ddot\chi\;;
\label{eqn:wUchi}
\end{equation}
and the 1PN solution to the Laplace equation (\ref{eqn:EFEwi}) for the DSX vector potential $w_j$ is 
\begin{equation}
w_j  =  \int  \frac{ \rho_o v_j} {|{\bm x} - {\bm x}'|} d^3 x'\;.
\label{eqn:gamma}
\end{equation}

The fluid's evolution is governed by rest-mass conservation, momentum conservation, and
energy conservation.  

The 1PN version of rest-mass conservation takes the following form:

\begin{subequations}
\begin{equation}
\rho_{*,t} + \bm\nabla\cdot (\rho_* \bm v) = 0 \;,
\label{eqn:masscons*}
\end{equation}
where
\begin{equation}
\rho_* = \rho_o u^0 \sqrt{-g} = \rho_o(1+\tfrac12 \bm v^2 + 3U)\; 
\label{eqn:rho*}
\end{equation}
\label{eqn:masscons}
\end{subequations}
(Eqs.\ (117) and (118) of Chandrasekhar \cite{ChandraPN}).  Here $u^0$ is the time component of the fluid's
4-velocity and $g$ is the determinant of the covariant components of the metric.

We shall discuss momentum conservation and energy conservation in the next two sections.

We note in passing that Chandrasekhar and many other researchers write their 1PN 
spacetime metric in a different gauge from our harmonic one.  The two gauges
are related by a change of time coordinate
\begin{subequations}
\begin{equation}
t_\textrm{C} = t_\textrm{H} - \tfrac12 \dot\chi\;,
\label{eqn:tCH}
\end{equation} 
and correspondingly the metric components in the two gauges are related by 
\begin{equation}
g_{00}^\textrm{C} = g_{00}^\textrm{H} + \ddot\chi\;, \quad
g_{0j}^\textrm{C} = g_{0j}^\textrm{H} + \tfrac12 \dot\chi_{,j}\;.
\label{eqn:gCH}
\end{equation}
\label{eqn:CH}
\end{subequations}
Here C refers to the Chandrasekhar gauge and H to our harmonic gauge.
DSX write their equations in forms that are invariant under the gauge
change (\ref{eqn:CH}).

\section{Momentum Density, Flux, and Conservation}
\label{sec:Momentum}

We now turn to our extension of the DSX formalism to include a Maxwell-like formulation of gravitational
momentum density, momentum flux, and momentum conservation.
Following Chandrasekhar \cite{ChandraPNcons}, Pati and Will \cite{PatiWill1} 
and others, we adopt the  Landau-Lifshitz pseudotensor as our tool 
for formulating these concepts.  

From Pati and Will's 2PN harmonic-gauge Eqs.\ (2.6), (4.4b) and (4.4c) for the pseudotensor, one can
deduce the following 1PN expressions for the 
the gravitational 
momentum density and momentum flux (stress) in terms of the
DSX gravitoelectric and gravitomagnetic fields $\bm g$ and $\bm H$:
\begin{subequations}
\begin{eqnarray}
(-g)t^{0j}_{\rm LL} \bm e_j &=& -\tfrac{1}{4\pi} \bm g \times \bm H + 
\tfrac{3}{4\pi} \dot U_N \bm g \;,
\nonumber \\
\label{eqn:LL0j} 
\\
(-g) t^{ij}_{\rm LL} &=& \tfrac{1}{4\pi}(g_i g_j - \tfrac12 \delta_{ij} g_k g_k)  \nonumber \\
&&+ \tfrac{1}{16\pi}(H_i H_j- \tfrac12 \delta_{ij} H_k H_k )- \tfrac{3}{8\pi} \dot U_N^2\delta_{ij}  \;. \nonumber \\
\label{eqn:LLij}
\end{eqnarray}
\label{eqn:LL}
\end{subequations}
Each equation is accurate up to corrections of order $U_N$ times 
the smallest term on the right side (2PN corrections).

For comparison, in flat spacetime the electromagnetic momentum density is $\tfrac{1}{4\pi} \bm E \times \bm B$ 
and the momentum flux is $\tfrac{1}{4\pi} (E_iE_j - \tfrac12 \delta_{ij} E_k E_k) + \tfrac{1}{4\pi} (B_iB_j - \tfrac12 \delta_{ij} B_k B_k)$. 
\textit{Aside from a sign in Eq.\ (\ref{eqn:LL0j}) and the two terms involving $\dot U_N$,   
the gravitational momentum flux and density (\ref{eqn:LL}) are identical to the electromagnetic ones
with $\bm E \rightarrow \bm g$ and $\bm B \rightarrow \bm H$.}  Therefore, by analogy with the 
electromagnetic case, there are {\it gravitational tensions}
$|\bm g|^2/8\pi$ and $|\bm H|^2/8\pi$ parallel to gravitoelectric and gravitomagnetic field lines, and
{\it gravitational pressures} of this same magnitude orthogonal to the field lines.  This makes
the gravitoelectric and gravitomagnetic fields $\bm g$ and $\bm H$ powerful
tools for building up physical intuition about the distribution and flow of gravitational momentum.  We
use them for that 
in our studies of compact binaries \cite{CKNT}, relying heavily on Eqs.\ (\ref{eqn:LL}). 

Here are some hints for deducing Eqs.\ (\ref{eqn:LL}) from Pati and Will \cite{PatiWill1} 
(henceforth PW): (i) Show that the last two terms in (2.6) of PW
are of 2PN order for $\{\alpha,\beta\} = \{0j\}$ or $\{ij\}$ and so can be ignored, whence
$16\pi (-g) t^{\alpha\beta}_{\rm LL} = \Lambda^{\alpha\beta}$.     
(ii) Show that our notation is related to that of PW by $\chi = - X$, $U$ the same, $w=U-\frac12 \ddot\chi
= \frac14(N+B) - \frac18(N+B)^2$ [for the last of these cf.\ PW (5.2), (5.4a,c)], and at Newtonian
order $U_N = \frac14 N$.
(iii) In PW (4.4b,c) for $\Lambda^{\alpha\beta}$,
keep only the Newtonian and 1PN terms: the first curly bracket in (4.4b) and first and second curly 
brackets in (4.4c).  Rearrange those terms so they involve only  $\bm K$, $N+B$ and the
Newtonian-order $N$, use
the above translation of notation and use the definitions (\ref{eqn:DSXghDef}) 
of $\bm g$ and $\bm H$.  
Thereby, bring 
PW (4.4b,c) into the form (\ref{eqn:LL}).

In the Landau-Lifshitz formalism, the local law of 4-momentum conservation ${T^{j\mu}}_{;\mu}=0$
takes the form 
\begin{equation}
[(-g) (T^{j\mu} + t^{j\mu}_\textrm{LL})]_{,\mu} = 0
\label{eqn:momconsdiv}
\end{equation} 
(Eqs.\ (20.23a) and (20.19) of \cite{MTW}, or (100.8) of \cite{LL62}).   Here (as usual), commas denote
partial derivatives, and semicolons denote covariant derivatives.   \emph{This is the conservation law that
we use in our studies of momentum flow in compact binaries} \cite{CKNT}.  

When dealing with material bodies (e.g.\ in DSX) rather than with the 
vacuum outside compact bodies, an alternative Maxwell-like version of 
momentum conservation is useful.  Specifically, 
using expressions (\ref{eqn:LL}) and
$(-g) = 1+4U_N$ [from Eq.\ (\ref{eqn:DSXmetric}) 
with $w=U_N$ at leading order], 
and using the field equations (\ref{eqn:gMaxwell})
for $\bm g$ and $\bm H$, the conservation law (\ref{eqn:momconsdiv}) can be rewritten in the following simple 
Lorentz-force-like form (Eq.\ (4.3) of Damour, Soffel and Xu's Paper II \cite{DSX2})
\begin{eqnarray}
&&[(1+4U_N)T^{i0}]_{,t} + [(1+4U_N) T^{ij}]_{,j}  
\label{eqn:momcons} \\
&&\quad = (T^{00} + T^{jj}) g_i + \epsilon_{ijk} T^{0j}H_k  \:. \nonumber
\end{eqnarray}
Here  the Levi-Civita tensor $\epsilon_{ijk}$  produces a cross product of the momentum density with
the gravitomagnetic field.  For
comparison, in flat spacetime, the momentum conservation law for a charged medium interacting with  electric and magnetic  fields $E_i$ and $B_i$ has the form ${T^{i0}}_{,t} + {T^{ij}}_{,j} = \rho_e E_i + \epsilon_{ijk} J_j B_k$, where
$\rho_e$ is the charge density and $J_j$ the charge flux (current density).
\textit{The right-hand side of Eq.\ (\ref{eqn:momcons}) (the gravitational force density) is identical
to that in the electromagnetic case, with $\rho_e \rightarrow (T^{00}+T^{jj})$, $J_j \rightarrow T^{0j}$,
$\bm E \rightarrow \bm g$, and $\bm B \rightarrow \bm H$.}  Again, this makes $\bm g$ and $\bm H$
powerful foundations for gravitational intuition.

For a perfect fluid, the components of the 1PN stress-energy tensor, which appear in the momentum conservation law
(\ref{eqn:momcons}), are (Eqs.\ (20) of Chandrasekhar \cite{ChandraPN})
\begin{subequations}
\begin{eqnarray}
T^{00}&=& \rho_o(1+\Pi + \bm v^2 + 2U_N) \;,
\label{eqn:T00}\\
T^{i0} &=& \rho_o(1+\Pi + v^2 + 2U_N + P/\rho_o) v_j\; ,
\label{eqn:Ti0}\\
T^{ij} &=& \rho_o (1+\Pi + \bm v^2 + 2U_N + P/\rho_o)v_i v_j \nonumber\\
&&+ P(1-2U_N) \delta_{ij} \;,
\label{eqn:Tij} \\
T^{00} + T^{jj} &=& \rho(1+\Pi + 2 \bm v^2+ 2U_N + 3P/\rho_o)\;.
\label{eqn:T00+Tjj}
\end{eqnarray}
\label{eqn:stress-energy}
\end{subequations}

\section{Energy Conservation}
\label{sec:energyconservation}

For a perfect fluid, the exact (not just 1PN) law of energy conservation, when combined with
mass conservation and momentum conservation,  reduces to the first 
law of thermodynamics $d\Pi/dt = - P d(1/\rho_o)/dt$; so whenever one needs to invoke
energy conservation, the first law is the simplest way to do so.  For this reason, and because
deriving the explicit form of 1PN energy conservation 
$[(-g)(T^{0\mu} + t^{0\mu}_{\rm LL})]_{,\mu} =0$ is a very complex and delicate task
(cf.\ Sec.\ VI of \cite{ChandraNutku}), we shall not write it down here.  

However, we {\it do} write down the Newtonian law of energy conservation in harmonic gauge, 
since we will occasionally need it in our future papers.
Chandrasekhar calculated $(-g)(T^{0\mu} + t^{0\mu}_{\rm LL})$ in \cite{ChandraPNcons};
his Eqs.\ (48) and (57) are the time-time and time-space components, respectively.
When one writes the expressions in terms of the ``conserved rest-mass density'' $\rho_*$ 
[Eq.\ (\ref {eqn:rho*})] and in our Maxwell-like form, Newtonian conservation of energy states that
\begin{eqnarray}
&&\left[\rho_* (1+\Pi + \frac12 v^2 + 3U_N) - \frac{7}{8\pi}\bm g\cdot \bm g \right]_{,t} \nonumber\\
&& + \bm \nabla \cdot \left[\rho_* \bm v(1+\Pi +\frac{P}{\rho}+ \frac12 v^2 + 3U_N) \right.\nonumber\\
&&\left.+\frac{3}{4\pi} \dot U_N \bm g - \frac{1}{4\pi} {\bm g}\times{\bm H} \right] = 0\;.
\label{eqn:EnConsPrelim}
\end{eqnarray}
While this equation is perfectly correct, it expresses Newtonian energy conservation in terms of the
post-Newtonian gravitomagnetic field ${\bm H}$.
It is possible to rewrite the $\bm H$-dependent term using the relationship,
$\nabla \cdot [-1/(4\pi) ({\bm g}\times{\bm H})] = \nabla\cdot[-4U_NT^{0j}{\bm e}_j + (1/\pi)U_N\dot{\bm g} ]$,
which is accurate up to corrections of order ${\bm g}\cdot \dot{\bm H}$.
This relationship can be found by applying Eq.\ (\ref{eqn:curlg}) once and (\ref{eqn:curlH}) twice,
in combination with elementary vector-calculus identities.
The statement of Newtonian energy conservation then depends only upon the Newtonian potential 
and its gradient and time derivative:
\begin{eqnarray}
&&\left[\rho_* (1+\Pi + \frac12 v^2 + 3U_N) - \frac{7}{8\pi}\bm g\cdot \bm g \right]_{,t} \nonumber\\
&& + \bm \nabla \cdot \left[\rho_* \bm v(1+\Pi +\frac{P}{\rho}+ \frac12 v^2 -U_N) \right. \nonumber\\
&& \left. +\frac{3}{4\pi} \dot U_N \bm g + \frac{1}{\pi} U_N \dot {\bm g} \right] = 0\;.
\label{eqn:EnCons}
\end{eqnarray}
Notice that going from Eq.\ (\ref{eqn:EnConsPrelim}) to Eq.\ (\ref{eqn:EnCons}) involves adding a divergence-free piece to the energy flux, so it entails changing how the energy flux is localized --- a change that strictly speaking takes the energy flux out of harmonic gauge.

If the coefficients of the gravitational terms in Eq.\ (\ref{eqn:EnCons}) look unfamiliar, it is because
even at Newtonian order, the density and flux of gravitational energy are
gauge-dependent.  In some other gauge, they will be different; see Box
12.3 of \cite{BlandfordThorne}.

\section{Gravitational Potentials in the Vacuum of a System of Compact, Spinning Bodies}
\label{sec:vacuum}

For a system of compact, spinning bodies (neutron stars or black holes),
the gravitational potentials $U_N$, $U$, $w_i$
and $\chi$ in the vacuum between the bodies take the following forms (in a slightly different harmonic
gauge than the one used in Sec.\ \ref{sec:PerfectFluid} for fluids): 
\begin{subequations}
\begin{eqnarray}
U_N &=& \sum_A  \frac{M_A}{r_A} \;, 
\label{eqn:UNpt} \\
U &=&  \sum_A  \frac{M_A}{r_A} \left( 1+ \tfrac32  \bm v_A^2 
- \sum_{B\ne A} \frac{M_B}{r_{AB}}\right) 
\nonumber\\
&& + 2 \sum_A \frac{\epsilon_{ijk} v_A^i S_A^j n_A^k}{{r_A}^2}\;,
\label{eqn:Upt} \\
\chi &=& - \sum_A M_A r_A\;,
\label{eqn:chipt} \\ 
w &=& U -\frac12\ddot\chi\;,
\label{eqn:wpt} \\
w_i &=&  \sum_A \frac{M_A v_A^i}{r_A} +\frac12 \sum_A \frac{\epsilon_{ijk} S_A^j n_A^k}{{r_A}^2}\;,
\label{eqn:gammapt}
\end{eqnarray}
\label{eqn:potentialspt}
\end{subequations}
Here the notation is that of Sec.\ IV of Thorne and Hartle \cite{ThorneHartle}: the sum is over
the compact bodies labeled by capital Latin letters $A, B$; $M_A$, $S_A^j$ and
$v_A^j$ are the mass, spin angular momentum and coordinate velocity of body $A$; 
$r_A$ is the coordinate distance from the field point to the center of mass 
of body $A$; $r_{AB}$ is the coordinate distance between the centers of mass of bodies $A$
and $B$;  $n_A^j$ is the unit vector
pointing from the center of mass of body $A$ to the field point; and $\epsilon_{ijk}$ is
the Levi-Civita tensor.  

Equations  (\ref{eqn:potentialspt}) for the potentials can be deduced by comparing our 1PN spacetime metric
coefficients 
[Eqs.\ (\ref{eqn:DSXmetric}), (\ref{eqn:wUchi})] 
with those in 
Eqs.\ (2.4), (5.11) and (5.14) of Racine and Flanagan \cite{RacineFlanagan}
or in
Eqs.\ (4.4) of Thorne and Hartle \cite{ThorneHartle}.\footnote{Racine and Flanagan
specialize DSX to a system of compact bodies with a complete set of nonzero multipole moments.
We neglect all moments except the bodies' masses and spins (see fifth paragraph of Sec.~\ref{sec:intro}).
Our notation is related to that of Racine and Flanagan by $U_N = -\Phi$, $w=U-\frac12 \ddot\chi = -(\Phi+ \psi), w_i=-\frac14\zeta_i$.  The Racine-Flanagan derivation of Eqs.\ (\ref{eqn:potentialspt}) avoids considering
the internal structures of the bodies and it therefore is directly valid for black holes.  The Thorne-Hartle
derivation relies on the pioneering analysis of Einstein, Infeld and Hoffman \cite{EIH} which uses a
fluid description of the bodies' interiors. Thorne and Hartle extend Eqs.\ 
(\ref{eqn:potentialspt})
to black holes by the equivalent of Damour's 
``effacement'' considerations, Sec.\ 6.4 of \cite{Damour300}.}


\section{Conclusion}
\label{sec:conclusion}

In our Maxwell-like formulation of the 1PN approximation to general relativity
for fluid bodies, the evolution of the fluid and gravitational fields is governed by: (i)  the law of momentum conservation
(\ref{eqn:momcons}), (\ref{eqn:stress-energy}) (which can be thought of as evolving the fluid velocity $v_j$); (ii) the law of mass
conservation (\ref{eqn:masscons})
(which can be thought of as evolving the mass density $\rho_o$); (iii) the equation of state $P(\rho_o)$ and
first law of thermodynamics $d\Pi = - P d(1/\rho_o)$ (which determine $P$ and $\Pi$ once $\rho_o$ is known);
Eqs.\ (\ref{eqn:UNU}), (\ref{eqn:gamma}), (\ref{eqn:chi}) for the gravitational potentials $U$, $\chi$, and $w_j$;  and Eqs.\ (\ref{eqn:DSXghDef}) 
or (\ref{eqn:gMaxwell}) for the
gravitoelectric and gravitomagnetic fields $\bm g$, $\bm H$.

When specialized to a system of compact bodies, e.g.\ a binary made of black holes or neutron stars, 
the system is governed by: (i) 1PN equations of motion and precession for the binary (not given in this paper; see, e.g., 
Eqs.\ (4.10), (4.11) and (4.14) of \cite{ThorneHartle});
(ii) momentum flow within the binary as described by the Landau-Lifshitz pseudotensor
(\ref{eqn:LL})  and its conservation law (\ref{eqn:momconsdiv}), in which the gravitoelectric and 
gravitomagnetic fields are 
expressed as sums over the bodies via Eqs.\ (\ref{eqn:potentialspt}) and (\ref{eqn:DSXghDef}); and (iii) other tools developed by Landau and Lifshitz (Sec.\ 100 of \cite{LL62}).
We are finding this formalism powerful in gaining insight into compact binaries \cite{CKNT}.

\section*{Acknowledgments}   For helpful discussions we thank Yanbei Chen
and Drew Keppel.  We also thank Luc Blanchet for bringing references \cite{BlanchetDamour} and \cite{DSX1} to our attention, and anonymous referees
for helpful critiques and suggestions.  
This research was supported in part by NSF grants  PHY-0601459
and PHY-0653653 and by a Caltech Feynman Fellowship to JDK.

\end{document}